\documentclass[10pt]{asme2ej}

\usepackage{epsfig} 
\usepackage{amssymb}
\usepackage{comment}
\usepackage{amsmath}
\usepackage{caption}
\usepackage{subcaption}
\usepackage[section]{placeins}
\usepackage{graphicx,epsfig}
\usepackage{units}
\usepackage{color}
\usepackage [sc]{mathpazo}

\definecolor{cadmiumgreen}{rgb}{0.0, 0.42, 0.24}

\title{Ice penetration by a bluff-body melting probe}

\author{Kerstin Weinberg\thanks{corresponding author}
    \affiliation{
	Department of Mechanical Engineering\\
	Universit{\"a}t Siegen, 57068 Siegen, Germany\\
    Email: kerstin.weinberg@uni-siegen.de
    }	
}

\author{Michael Ortiz
    \affiliation{Division of Engineering and Applied Science\\
     California Institute of Technology,
     Pasadena, CA 91125, USA\\
     Email: ortiz@caltech.edu
    }
}

\begin{document}

\maketitle

\begin{abstract}
We analyze the operation of melting probes as a Stefan problem for the liquid/solid interface surrounding the probe. We assume that the liquid layer is thin and, therefore, amenable to analysis by lubrication theory. The resulting Stefan problem is solvable in closed form. The solution determines the dependence of the penetration speed on the temperature differential between the probe and the surrounding ice, the size, shape and weight of the probe, the viscosity of liquid water and the thermal properties of solid ice.
\end{abstract}


\section{Introduction}

\smallskip

{Melting probes} are hot probes that melt the surrounding ice and descend through it under the action of gravity. The probe carries its own heat source and can operate automatically in the absence of controls. The idea to use melting probes for exploring the subsurface layers of planetary ice sheets goes back, at least, to the pioneering work of Karl Philberth \cite{Philberth1962}, who constructed such probes and performed field tests in Greenland. Since then, a number of programs and studies have developed the concept further, cf.  \cite{Komle2018} for a literature review. The use of melting probes for penetrating thick icy layers is particulary appealing in places where conventional drilling is infeasible, e.~g., in extraterrestrial icy habitats on Mars and on the moons Europa, Enceladus and Titan, cf., e.~g., \cite{Chyba2001, Marion2003, Powell2005, Hand2009, Bulat2011}, or in isolated terrestrial cryoenvironments, such as permafrost cryopegs and subglacial ecosystems, cf., e.~g., \cite{Murrayw2012}.

The conventional analysis of melting probes is based on thermal analysis and balance of energy. Typically, three processes are considered in the analysis \cite{Ulamec2007}: i) heating of the ice surrounding the probe to its melting temperature; ii) melting the solid ice to liquid water; and iii) conductive and radiative losses. The power required for the first two processes can be estimated simply in terms of the heat capacity of ice and the latent heat of melting. The determination of conductive losses requires the solution of a transient or steady-state heat conduction problem for the exterior domain of the probe. The speed of descent of the probe is then estimated through a balance of energy between heat supply and the latent heat required to melt the ice. However, in a more comprehensive description of the process the speed of descent results from a competition between gravity, which pushes the probe down, viscous dissipation in the liquid layer that surrounds the probe, latent heat and radiative losses. The conventional thermal analysis thus neglects potentially important parameters such as the weight of the probe and the viscosity of liquid water.

In the present work, we analyze the penetration of melting probes as a {Stefan problem} for the liquid/solid interface surrounding the probe. We start with an analysis the  temperature field of a moving probe. The liquid layer around the probe  is assumed to be thin and, therefore, amenable to analysis by lubrication theory. The resulting Stefan problem is solvable in closed form within the lubrication theory approximation. The solution determines the dependence of the penetration speed on the temperature differential between the probe and the surrounding ice, the size, shape and weight of the probe, the viscosity of liquid water and the thermal properties of solid ice.

\section{Supporting experiments}\label{H8Dvzt}

\smallskip

\begin{figure}[h]\centering
\includegraphics[width=0.8\linewidth]{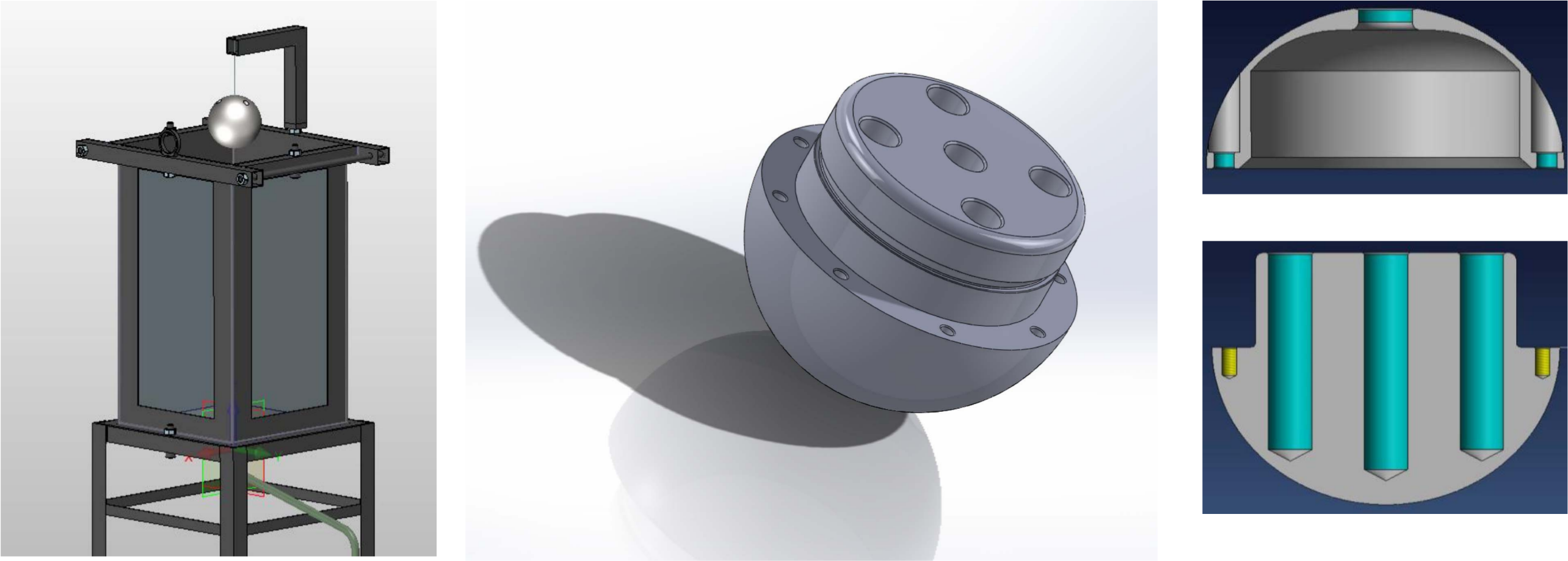}
	\caption{Experimental setup: Rack with an ice block (left); lower half of the spherical penetrator (middle); detail of the heating cartridges (right).} \label{eisVersuchCAD}
\end{figure}

We begin by demonstrating the melting probe concept by means of a simple supporting experiment. The setup consists of a water ice block in a container of size $0.5\times0.5\times1\,$m, Fig.~\ref{eisVersuchCAD} (left). The walls of the container are made of transparent PMMA plates, which allows the evaluation of the ice quality, and are affixed to a steel frame. The box is filled with water and cooled down slowly to $-18^\circ$C in order to obtain an ice block of good quality, i.~e., without layers or large bubbles. Once frozen, the block is insolated with polystyrene panels.

\begin{figure}[t]\centering
\includegraphics[width=0.45\linewidth]{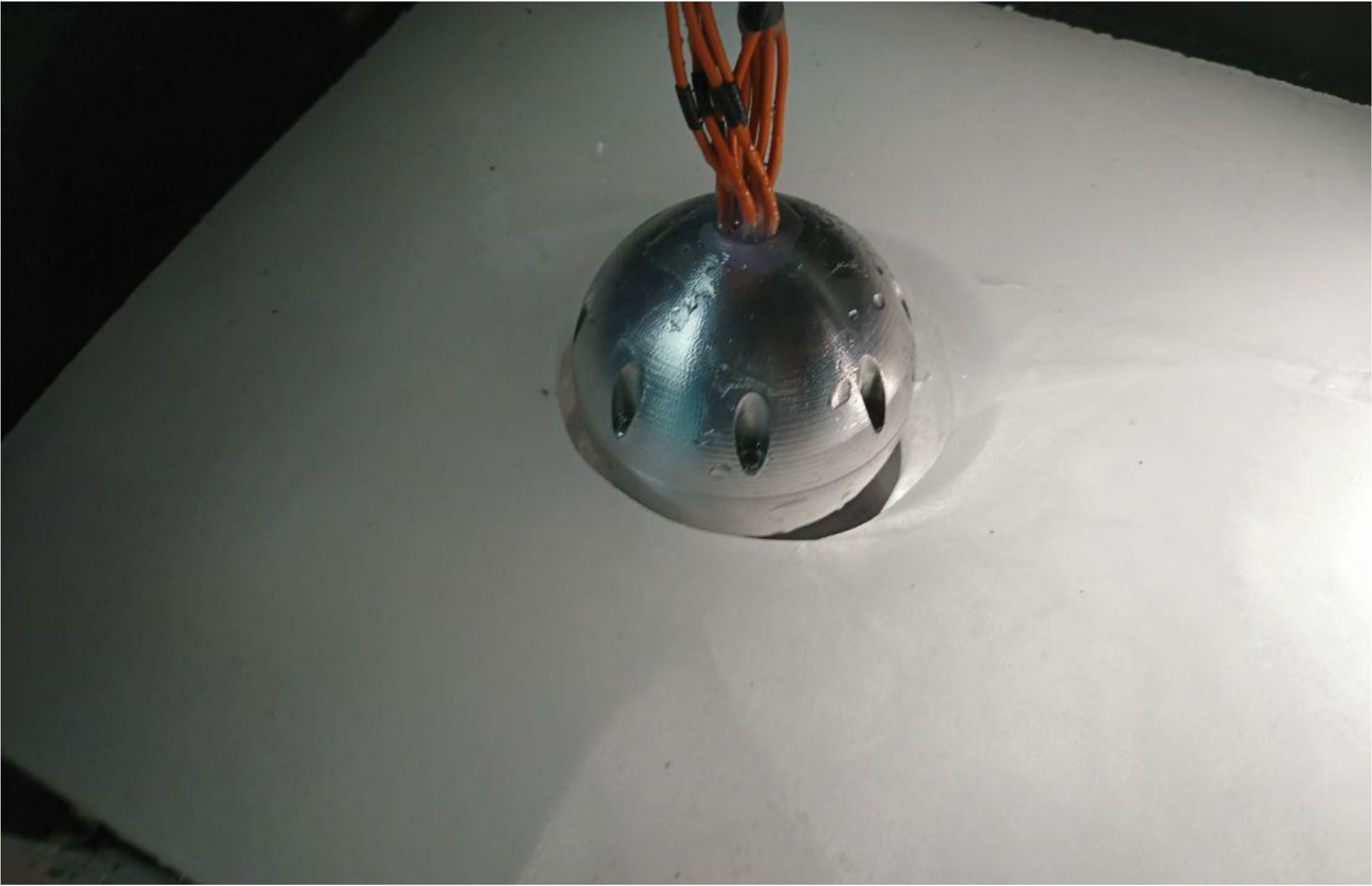}
	\caption{Snapshot of the experiment as the heated probe starts to melt the ice.} \label{eisVersuchFoto}
\end{figure}

The probe is spherical of outer diameter $80$\,mm, is made of an aluminum (AlCuMgPb) alloy with high thermal conductivity and has a weight $W = 7.24$ N.
The probe surface is treated by milling, drilling, scrubbing and a final surface polishing to achieve a smooth appearance. The probe consists of two identical halves, Fig.~\ref{eisVersuchCAD} (middle and right), and carries five heat cartridges. The cartridges are cylindrical (of size $10$\,mm $\times$ $50$\,mm) and supply a maximum nominal power of $180\,$W each. They are connected to an external electric circuit that controls the power supply. In the experiment the hot probe is  positioned on the top of the ice block. Driven by its own weight, it then moves through the ice, Fig.~\ref{eisVersuchFoto}. In order to avoid friction with the electrical cord, the water is evacuated as is melts.

\begin{table}\centering
\begin{tabular}{|c|c|c|c|c|c|}
    \hline
    $c_{\rm ice}$  &
    $\kappa_{\rm ice}$ &
    $c_{\rm water}$ &
    $\kappa_{\rm water}$  &
    $L$ &
    $\eta$
    \\
    $[{\rm J}/{\rm K}{\rm cm}^3]$ &
    $[{\rm W}/ {\rm K}{\rm m}]$ &
    $[{\rm J}/{\rm K}{\rm cm}^3]$ &
    $[{\rm W}/{\rm K}{\rm m} ]$ &
    $[{\rm J}/{\rm cm}^3]$ &
    $[{\rm mN\,s}/{\rm m}^2]$
    \\ \hline
    2.04 & 2.22 & 4.18 & 0.608 &    335 & 0.889
    \\ \hline
\end{tabular}
	\caption{Reference material constants for ice and water \cite{BauerBenensonB,Horvath1975}: $c -$ heat capacity per unit volume; $\kappa -$ thermal conductivity; $L -$ latent heat of fusion per unit volume, $\eta -$ viscosity} \label{tab:material}
\end{table}

We estimate the heat supply requirements by a simple energy balance
\begin{align}\label{heatQexperiment}
    Q = \left(c_{\rm ice} ( T_m - T_\infty) +  L \right) \!\cdot\!{\rm Vol}
\end{align}
where $T_\infty = -18^\circ$C is the ice temperature, $T_m = 0^\circ$C is the melting temperature, $c_{\rm ice}=\varrho c_p$ is the heat capacity per unit volume of ice with mass density $\varrho$, $L$ is the latent heat of melting per unit volume and ${\rm Vol}$ is the total volume of ice to melt. Reference values of these and related properties are collected in Table~\ref{tab:material}. Inserting the values into (\ref{heatQexperiment}) and taking the dimensions of the probe and containing box into account gives an estimate of $Q=934 \,$J. If we further assume an energy loss of $\nicefrac13$ and a duration of the experiment of about $20-30\,$min, the required power supply is estimated as $\dot{Q}=900$W. We thus verify that the heating cartridges located in the probe indeed have ample capacity to supply the requisite power over the duration of the experiment.

\begin{figure}[hb]\centering
\includegraphics[width=0.9\linewidth]{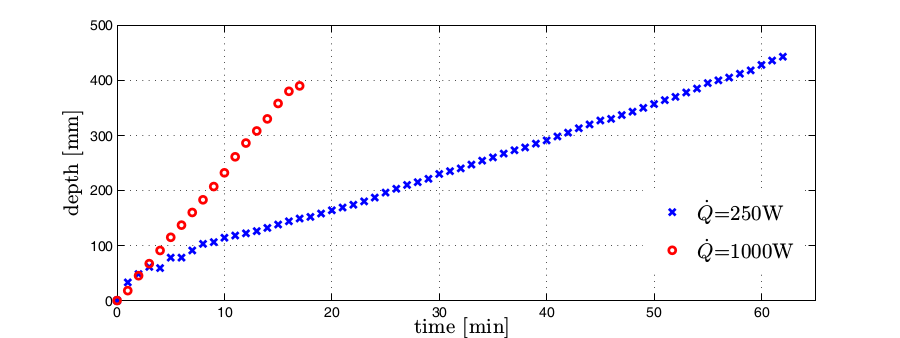}
	\caption{Measured depth {\sl vs.} time for two experiments.} \label{D2D5vu}
\end{figure}

Experiments were performed at two different power levels, $\dot{Q}=250\,$W and $\dot{Q}=1,000\,$W. Fig.~\ref{D2D5vu} shows the penetration depth as a function of time for the two cases. As may be seen from the figure, following an initial transient the penetration depth is a linear function of time, which is in turn indicative of a steady-state speed of descent. The measured speeds are: i)  $V=6.9\,$mm/min for $\dot{Q}=250\,$W; and ii) $V=23.9\,$mm/min for $\dot{Q}=1,000\,$W. These data correspond to the averaged values of \cite{Komle2018}; they suggest increasing power levels are required to sustain increasing speeds of descent.

The simple supporting experiment demonstrates the operation of melting probes and suggests  a steady state of constant speed of descent of the probe as well as a dependence of the power supply on the speed. A number of other relations, such as the dependence of the power supply and speed of descent on the size and shape of the probe and on material properties such heat capacity, latent heat, thermal conductivity and viscosity remain to be elucidated. The analysis that follows is aimed at addressing these questions.

\section{Thermal analysis}

\smallskip

\begin{figure} [h]\centering
\includegraphics[width=0.55\linewidth]{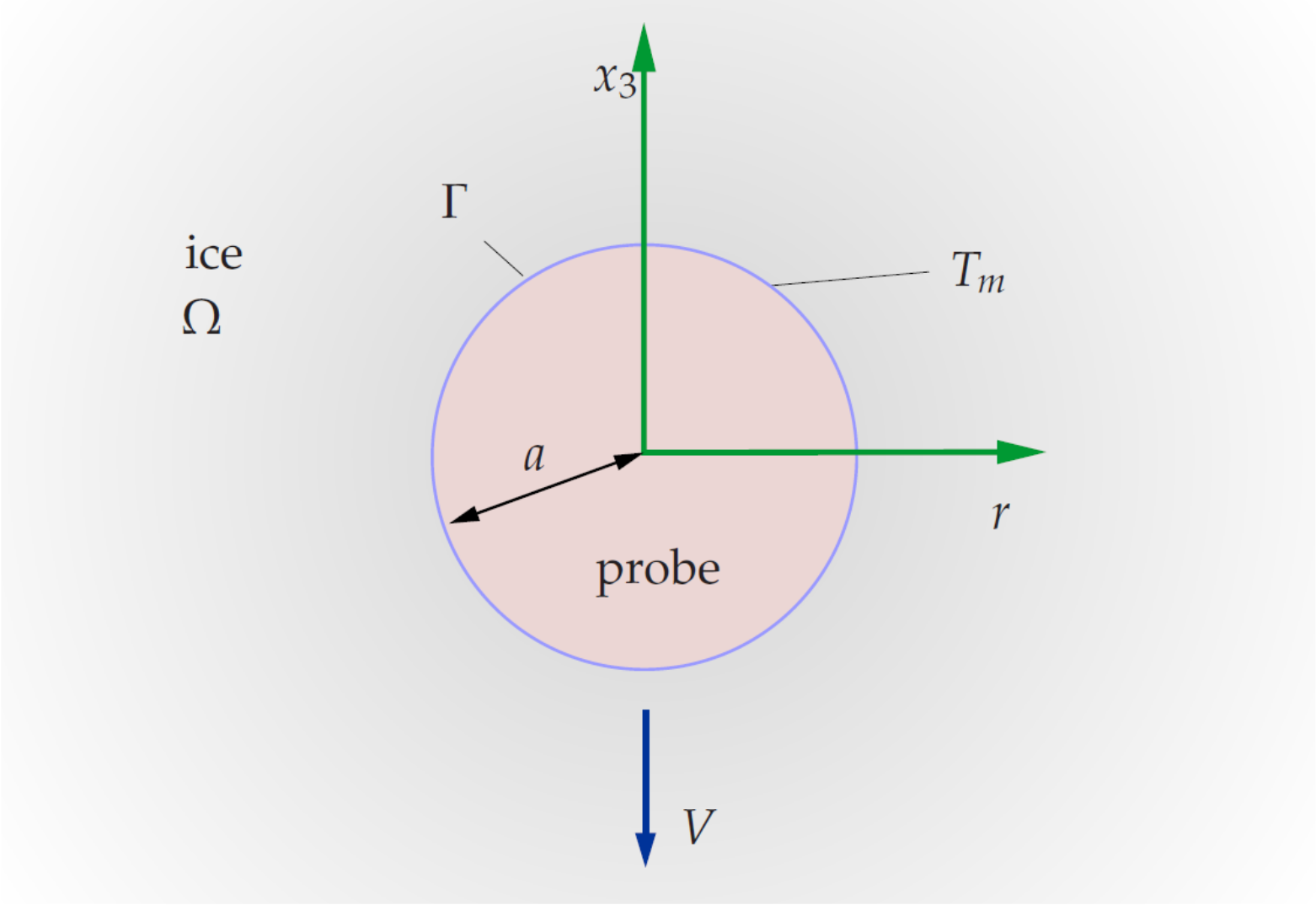}
	\caption{Melting probe system for thermal analysis.} \label{europaTemp}
\end{figure}

We begin by determining the dependence of the radiated heat rate on the sinking speed. We assume that the probe is surrounded by a layer of liquid water of thickness much smaller than the dimensions of the probe. In addition, the interface between the liquid layer and the surrounding solid ice is at the melting temperature $T_m$. Under these assumptions, the temperature field $T(x,t)$ in the solid ice is determined by the problem
\begin{subequations}
\begin{align}
    &
    c \frac{\partial T}{\partial t}
    =
    \kappa \Delta T ,
    &&
    \text{in } \Omega(t) ,
    \\ &
    T = T_m ,
    &&
    \text{on } \Gamma(t) ,
    \\ &
    T = T_\infty ,
    &&
    r \to + \infty ,
\end{align}
\end{subequations}
where $t$ denotes time, $\Delta$ is the Laplacian operator, $\Omega(t)$ is the domain occupied by the solid ice at time $t$, $\Gamma(t)$ is the boundary of $\Omega(t)$ and the interface between the probe and solid ice, $c$ is the heat capacity per unit volume of ice, $\kappa$ is the thermal conductivity of ice, $T_m$ is the melting temperature of ice and $T_\infty$ is the background ice temperature. For simplicity, we take all material properties to be constant.

We further assume a steady state wherein the probe descends through the solid ice at constant speed $V$. To facilitate analysis of this steady state, we introduce a reference frame moving with the probe. In this moving frame, the governing equations become
\begin{subequations}
\begin{align}
    & \label{qIufr4}
    c  V \frac{\partial T}{\partial x_3}
    +
    {\kappa} \Delta T
    =
    0 ,
    &&
    \text{in } \Omega ,
    \\ &
    T = T_m ,
    &&
    \text{on } \Gamma ,
    \\ &
    T = T_\infty ,
    &&
    r \to + \infty ,
\end{align}
\end{subequations}
where $\Omega$ is the exterior domain of the probe in an infinite body of ice,  $\Gamma$ is its boundary and $x_3$ is the direction of descent.

\subsection{Spherical probe}

\smallskip

The case of a spherical probe, such as used in the supporting experiments reported in Section~\ref{H8Dvzt}, can conveniently be solved analytically in closed form. In this case, $\Omega = \{ r \geq a \}$ and $\Gamma = \{ r = a \}$, where $r$ is the radial distance to the origin and $a$ is the radius of the probe. Introducing spherical coordinates
\begin{equation}
    x_1 = r \sin\theta \cos\varphi ,
    \quad
    x_2 = r \sin\theta \sin\varphi ,
    \quad
    x_3 = r \cos\theta ,
\end{equation}
and the change of variables
\begin{equation}\label{dO5rle}
    T
    =
    T_\infty
    +
    \tilde{T} {\rm e}^{- \lambda x_3}
    =
    T_\infty
    +
    \tilde{T} {\rm e}^{- \lambda r \cos\theta} ,
\end{equation}
eq.~(\ref{qIufr4}) becomes
\begin{equation}\label{Dievl7}
    \lambda^2 \tilde{T}
    =
    \frac{1}{r^2}
    \frac{\partial}{\partial r}
    \left( r^2 \frac{\partial \tilde{T}}{\partial r} \right)
    +
    \frac{1}{r^2 \sin\theta}
    \frac{\partial}{\partial\theta}
    \left( \sin\theta \frac{\partial \tilde{T}}{\partial\theta} \right) ,
\end{equation}
where
\begin{equation}\label{9iekiE}
    \lambda
    =
    \frac{c  V}{2 {\kappa}}
\end{equation}
is a normalized speed of descent. Introducing the representation
\begin{equation}
    \tilde{T}(r,\theta)
    =
    R(r) S(\theta) ,
\end{equation}
eq.~(\ref{Dievl7}) further separates into the ordinary differential equations
\begin{subequations}
\begin{align}
    &
    \frac{1}{R}
    \frac{d}{dr}
    \left( r^2 \frac{dR}{dr} \right)
    -
    \lambda^2 r^2
    =
    n (n + 1) ,
    \\ &
    -
    \frac{1}{S}
    \frac{1}{\sin\theta}
    \frac{d}{d\theta}
    \left( \sin\theta \frac{dS}{d\theta} \right)
    =
    n (n + 1) ,
\end{align}
\end{subequations}
with $n \geq 0$. The solutions of these equations are
\begin{subequations}
\begin{align}
    &
    R_n(r)
    =
    A_n i_n(\lambda r) + B_n k_n(\lambda r) ,
    \\ &
    S_n(\theta)
    =
    C_n P_n(\cos\theta) + D_n Q_n(\cos\theta) ,
\end{align}
\end{subequations}
where $i_n(x)$ and $k_n(x)$ are the modified spherical Bessel functions of the first and second kind and $P_n(x)$, $Q_n(x)$ are the Legendre polynomials of the first and second kind, reqectively. Boundedness at $r \to +\infty$ requires $A_n = 0$, whereas boundedness at $\theta = \pm \pi/2$ requires $D_n = 0$. Under these conditions, the general solution takes the form
\begin{equation}\label{jL3xlA}
    \tilde{T}(r,\theta)
    =
    \sum_{n=0}^\infty
        T_{n} k_{n}(\lambda r) P_{n}(\cos\theta) .
\end{equation}
At the boundary $r=a$, we must have
\begin{equation}
    \tilde{T}(a,\theta)
    =
    \sum_{n=0}^\infty
        T_{n} k_{n}(\lambda a) P_{n}(\cos\theta)
    =
    (T_m-T_\infty) {\rm e}^{\lambda a \cos\theta},
\end{equation}
By the orthogonality of the Legendre polynomials, we have
\begin{equation}
   T_{n}
   =
   \frac{2 n + 1}{2 k_{n}(\lambda a)}
   (T_m-T_\infty)
   \int_{-1}^{1}
        {\rm e}^{\lambda a x}
        P_{n}(x)
   \, dx .
\end{equation}
An appeal to Rodrigues representation,
\begin{equation}
    P_n(x)
    =
    \frac{1}{2^n n!}
    \frac{d^n}{dx^n} (x^2-1)^n ,
\end{equation}
further gives
\begin{equation}
   T_{n}
   =
   \frac{2 n + 1}{2 k_{n}(\lambda a)}
   (T_m-T_\infty)
   \int_{-1}^{1}
        {\rm e}^{\lambda a x}
        \frac{1}{2^n n!}
        \frac{d^n}{dx^n} (x^2-1)^n
   \, dx .
\end{equation}
Integrating by parts $n$ times, we obtain
\begin{equation}
    T_{n}
    =
    \frac{2 n + 1}{2 k_{n}(\lambda a)}
    (T_m-T_\infty)
    \int_{-1}^{1}
        \frac{(-\lambda a)^n}{2^n n!}
        {\rm e}^{\lambda a x}
        (x^2-1)^n
    \, dx ,
\end{equation}
which evaluates to
\begin{equation}
   T_{n}
   =
   \Big( n + \frac{1}{2} \Big)
   \frac{\sqrt{\pi }}{k_{n}(\lambda a)}
   \left(\frac{\lambda a}{2}\right)^n
   \, _0\tilde{F}_1\left(\frac{3}{2} + n, \frac{\lambda^2 a^2}{4}\right)
   \,
   (T_m-T_\infty)
\end{equation}
where $_0\tilde{F}_1(a,z)$ is the regularized hypergeometric function $_0F_1(a,z)/\Gamma(a)$ and $\Gamma(z)$ is the Euler gamma function. The corresponding heat flux into the solid ice at $r=a$ is
\begin{equation}
    q(\theta)
    =
    -
    {\kappa} \frac{\partial T}{\partial r}
    =
    -
    {\kappa}
    \Big(
        \frac{\partial\tilde{T}}{\partial r}
        -
        \lambda \cos \theta \,
        \tilde{T}
    \Big)
    {\rm e}^{-\lambda a \cos\theta} ,
\end{equation}
which coincides with the Dirichlet-to-Neumann map for the exterior of the spherical domain specialized to a constant surface temperature $T_m$.

We expect the temperature far-field to tend asymptotically to that of a point heat source moving through the solid ice at the speed of the probe. We recall that the steady-state solution for a moving point heat source is \cite{CarslawJaeger1986}
\begin{equation}\label{gL3zou}
    T
    =
    T_\infty
    +
    \frac{C}{4 \pi r}
    {\rm e}^{-\lambda r (1 + \cos\theta)} ,
\end{equation}
with $\lambda$ as in (\ref{9iekiE}) and $C > 0$ quantifying the strength of the source. A straightforward calculation shows that (\ref{gL3zou}) indeed satisfies (\ref{qIufr4}). We determine $C$ by matching (\ref{gL3zou}) asymptotically to the full solution (\ref{dO5rle}) and (\ref{jL3xlA}). To leading order as $r \to +\infty$, we have
\begin{equation}\label{PoE3hi}
    T
    \sim
    T_\infty
    +
    {\rm e}^{- \lambda r \cos\theta}
        A_{0} k_{0}(\lambda r) P_{0}(\cos\theta) ,
\end{equation}
with
\begin{equation}
    A_0
    =
    (\lambda a)^2 (T_m - T_\infty) (1 + \coth(\lambda a)) ,
    \quad
    k_{0}(\lambda r)
    =
    \frac{{\rm e}^{-\lambda r}}{\lambda r} ,
    \quad
    P_0(x) = 1 ,
\end{equation}
whereupon (\ref{PoE3hi}) evaluates to
\begin{equation}\label{flA1Ro}
    T
    \sim
    T_\infty
    +
    (\lambda a)^2 (T_m - T_\infty) (1 + \coth(\lambda a))
    \frac{{\rm e}^{-\lambda r(1+\cos\theta)}}{\lambda r} .
\end{equation}
Matching constants between (\ref{gL3zou}) and (\ref{flA1Ro}), we find
\begin{equation}\label{aD3n81}
    C
    =
    4 \pi
    \lambda a^2 (T_m - T_\infty) (1 + \coth(\lambda a)) ,
\end{equation}
which fully characterizes the temperature far-field.

\begin{figure}[ht]\centering
\includegraphics[width=0.75\linewidth]{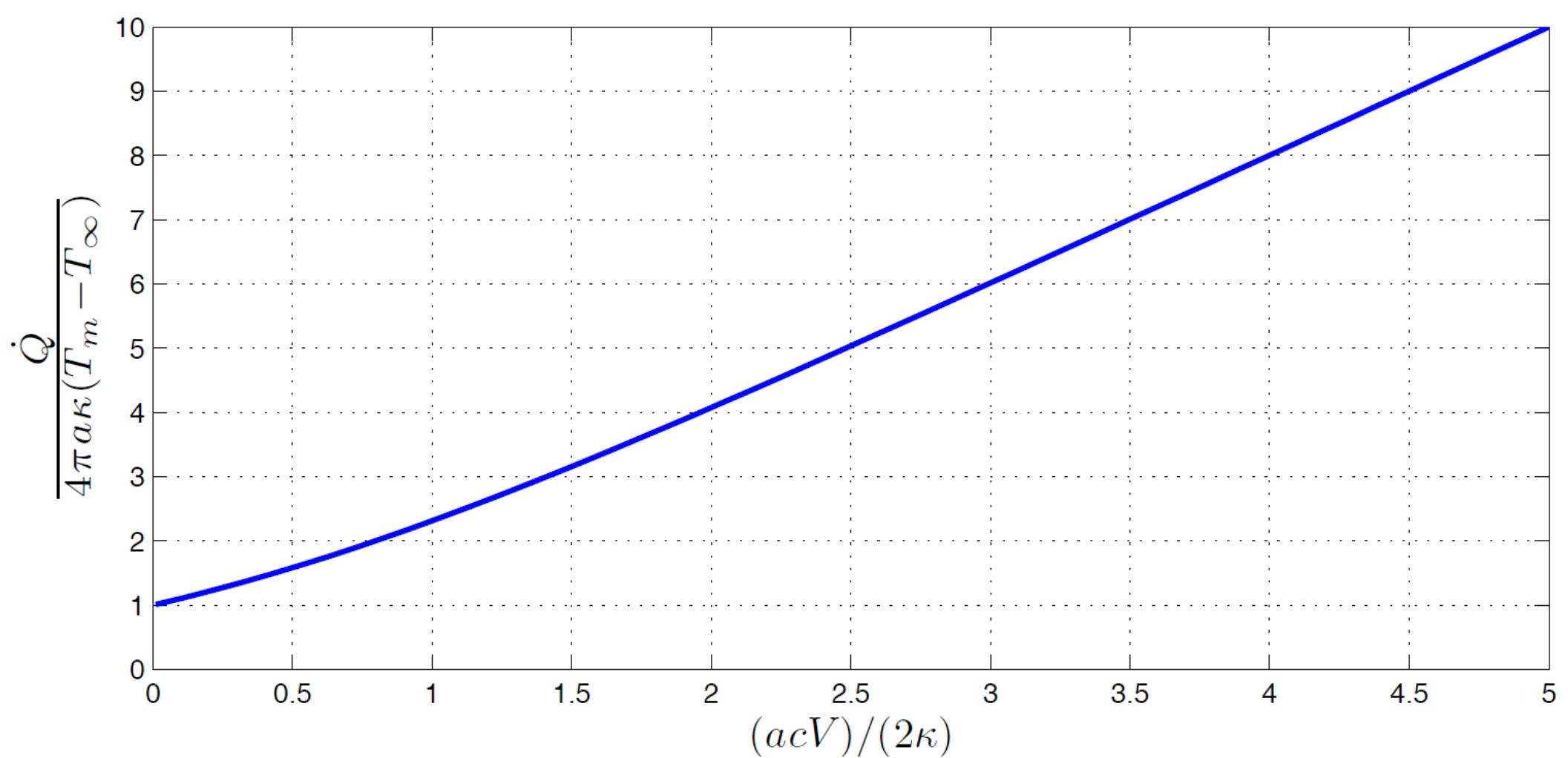}
    \caption{Normalized radiated heat rate {\sl vs}. speed of descent.} \label{CIeg1u}
\end{figure}

We may now compute the total rate of heat radiated from the probe as
\begin{equation}
    \dot{Q}_\infty
    =
    -
    \lim_{r\to \infty}
    \int_0^{\pi}
    \Big(
        c V T
        +
        {\kappa} \frac{\partial T}{\partial r}
    \Big)
        2\pi r^2 \sin\theta
    \, d\theta
    =
    C {\kappa} ,
\end{equation}
which, in view of (\ref{aD3n81}),  evaluates to
\begin{equation}\label{f1yV3X}
    \dot{Q}_\infty
    =
    4 \pi \kappa
    \lambda a^2 (T_m - T_\infty) (1 + \coth(\lambda a)) .
\end{equation}
This identity supplies the sought relation between the speed of descent $V$ and the radiated heat rate $\dot{Q}_\infty$ from the probe, Fig.~\ref{CIeg1u}. We note that the radiated heat rate increases with increasing speed of descent $V$, the dependence being approximately linear for large $V$,
\begin{equation}\label{heatSuppylinear}
    \dot{Q}_\infty
    =
    4 \pi a^2  c V (T_m - T_\infty) .
\end{equation}
We also see from this expression that, in the large $V$ regime the radiated heat rate scales with the square of the probe radius.

For the supporting experiment, with $T_m - T_\infty=18\,$K and a spherical probe of radius $a=4\,$cm and a velocity of $V=7\,$mm/min, the radiated evaluates to $\dot{Q}_\infty= 88\,$W, or about $\nicefrac{1}{3}$ of the actual power supply. This gap suggests the operation of dissipation mechanisms other than radiative losses. We investigate one possible such mechanism, viscous dissipation in the liquid layer, next.

\section{Lubrication analysis of the Stefan problem}

\smallskip

\begin{figure}[ht]\centering
\includegraphics[width=0.7\linewidth]{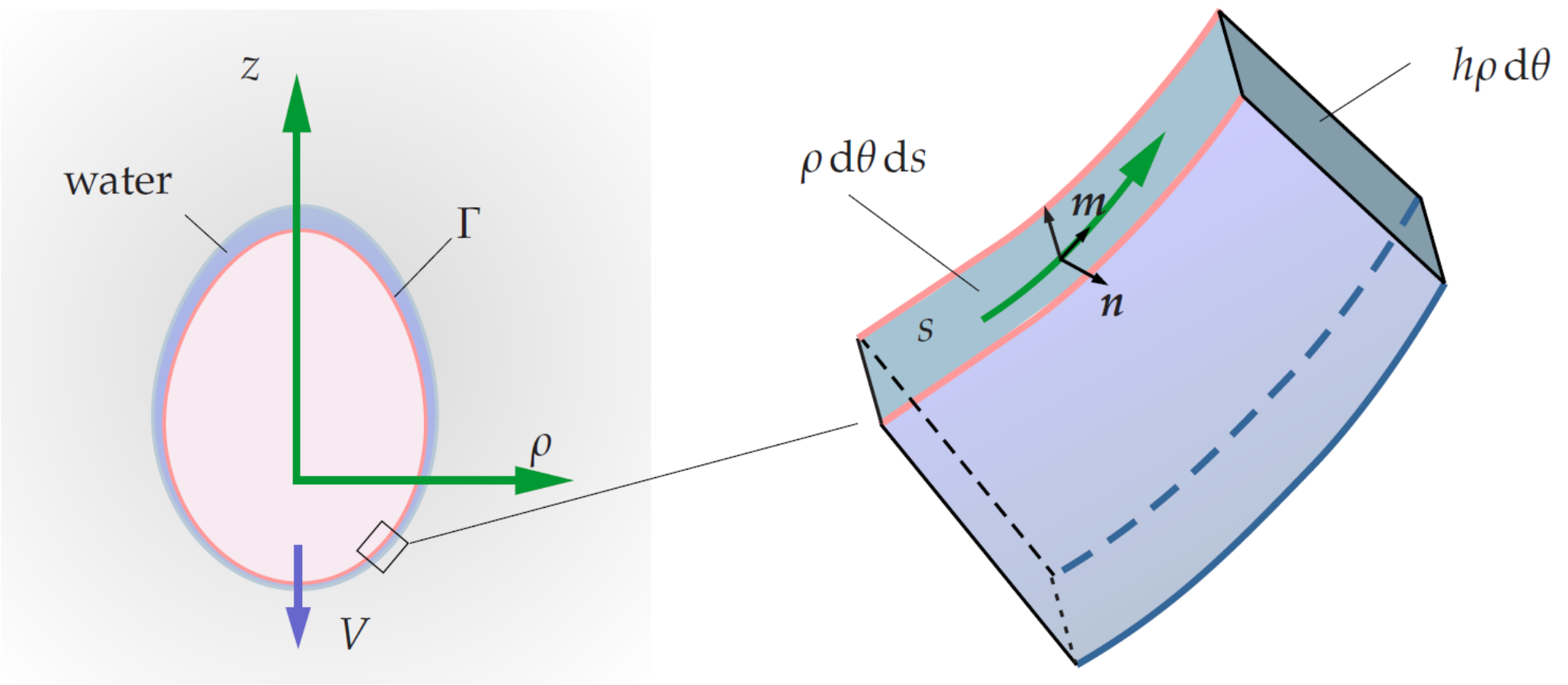}
\caption{Schematic of probe and liquid layer.} \label{europaSchnitt}
\end{figure}

We analyze the effect of the layer of liquid water surrounding the probe by recourse to a lubrication approximation, Fig.~\ref{europaSchnitt}. To this end, we introduce a system of cylindrical coordinates $\{\rho, \theta, z\}$ aligned with the axis of descent. We further assume cylindrical symmetry and describe the boundary $\Gamma$ of the probe by means parametric equations $\rho(s)$ and $z(s) $ in terms of the arc length $s$ measured from the axis. The unit tangent and outward normals are
\begin{equation}\label{ZDQuch}
    m(s) = (\rho'(s), z'(s))^T ,
    \qquad
    n(s) = (z'(s), - \rho'(s))^T ,
\end{equation}
where the prime denotes the derivative with respect to the arc-length $s$. In this parametrization, we have the identity
\begin{equation}\label{yz2N9S}
    \rho'^2(s) + z'^2(s) = 1 .
\end{equation}
We assume throughout steady state conditions. We note that the geometry of the liquid/solid interface, or, equivalently, the thickness of the liquid layer, is not known {\sl a priori} and is an outcome of the analysis, in the spirit of Stefan problems. The main objective of the analysis is to determine the extent of viscous dissipation in the liquid layer as a function of the speed of descent.

\subsection{Mass conservation}

\smallskip

Conservation of mass in a control volume $dV = \rho h  \, d\theta \, ds$ coincident with the liquid layer gives
\begin{equation}
\begin{split}
    &
    (-V e_z) \cdot n(s) \, ( \rho(s) \, d\theta \, ds )
    = 
    (v(s) + v'(s) \, ds))(h(s) + h'(s)) (\rho(s) + \rho'(s) ds)\, d\theta
    - 
    v(s) h(s) \rho(s) \, d\theta ,
\end{split}
\end{equation}
where $v(s)$ is the mean velocity within the liquid layer and $h(s)$ is its thickness. Simplifying, we obtain the relation
\begin{equation}
    V \rho(s) \rho'(s)
    =
    \frac{d}{ds} (\rho(s) h(s) v(s)) .
\end{equation}
With $s$ measured from the axis, integration with respect to $s$ gives
\begin{equation}
    \frac{V}{2} \rho(s)
    =
    h(s) v(s) ,
\end{equation}
whence the mean velocity of the liquid water follows as
\begin{equation}\label{4leWri}
    v(s)
    =
    \frac{V}{2} \frac{\rho(s)}{h(s)} .
\end{equation}

\subsection{Weight-drag equilibrium}

\smallskip

The descent of the melting probe is driven by gravity.
We assume static equilibrium between the weight $W$ of the probe, its buoyancy ${\varrho g} \Omega$ and the total drag force $F_d$ resulting from the viscosity of the fluid layer, i.~e.,
\begin{equation}\label{3r6UfT}
    F_d = W - {\varrho g} \Omega \geq 0,
\end{equation}
where ${\varrho g}$ is the mass density of water, $\Omega$ is the henceforth volume displaced by the probe and we assume that the probe is heavier than water. In requiring this equilibrium, we implicitly regard point-load reactions, such as would occur at a point-contact between the bottom of the probe and the ice, as unphysical. The total drag force on the probe is
\begin{equation}\label{2jTFDp}
    F_d
    =
    \int_\Gamma
        \tau(s) m_z(s)
        2 \pi \rho(s)
    \, ds ,
\end{equation}
where $\tau(s)$ is the shear stress on the probe  and the integral is effected over a probe meridian. Further assuming a linear velocity profile within the fluid layer, corresponding to local Couette flow, and using (\ref{ZDQuch}), eq.~(\ref{2jTFDp}) becomes
\begin{equation}
    F_d
    =
    \int_\Gamma
        \eta \frac{2v(s)}{h(s)} z'(s)
        2 \pi \rho(s)
    \, ds .
\end{equation}
Finally, inserting (\ref{4leWri}) we obtain
\begin{equation}\label{HuNyrU}
    F_d
    =
    2 \pi \eta  V
    \int_\Gamma
        \frac{\rho^2(s)z'(s)}{h^2(s)}
    \, ds ,
\end{equation}
which scales linearly with the speed of descent for fixed fluid layer thickness.

\subsection{Viscous dissipation}

\smallskip

Assuming again local Couette flow, the total dissipation evaluates to
\begin{equation}
    D
    =
    \int_\Gamma
        \eta
        \Big( \frac{2 v(s)}{h(s)} \Big)^2
        h(s) 2 \pi \rho(s)
    \, ds .
\end{equation}
Alternatively, inserting relation (\ref{4leWri}) we obtain
\begin{equation}\label{23HK55}
    D
    =
    2 \pi \eta V^2
    \int_\Gamma
        \frac{\rho^3(s)}{h^3(s)}
    \, ds ,
\end{equation}
which, as expected, scales with the square of the speed of descent for fixed fluid layer thickness.

\subsection{Optimal fluid layer profile}

\smallskip

It remains to determine the fluid layer profile $h(s)$. We note that $h(s)$ defines the configuration of the fluid layer and, therefore, it may be expected to be the result of configurational equilibrium. Specifically, we assume that $h(s)$ accommodates itself so as to minimize the total viscous dissipation (\ref{23HK55}) subject to the equilibrium constraint (\ref{2jTFDp}). Enforcing this constraint by means of a Lagrange multiplier $\mu$, we obtain the Lagrangian,
\begin{equation}
    S
    =
    2 \pi \eta V^2
    \int_\Gamma
        \frac{\rho^3(s)}{h^3(s)}
    \, ds
    -
    \mu
    \Big(
        2 \pi \eta  V
        \int_\Gamma
            \frac{\rho^2(s)z'(s)}{h^2(s)}
        \, ds
    \Big) .
\end{equation}
Stationary of $S$ with respect to $h(s)$ gives
\begin{equation}\label{hPQPpg}
    h(s)
    =
    \frac{3 V}{2 \mu}
    \frac{\rho(s)}{z'(s)} ,
\end{equation}
which, inserted into (\ref{HuNyrU}), gives
\begin{equation}
    F_d
    =
    \frac{8 \pi \eta \ell}{9 V}
    \mu^2 ,
\end{equation}
with
\begin{equation}
    \ell
    =
    \int_\Gamma
        z'^3(s)
    \, ds
\end{equation}
defining an effective probe size. The equilibrium condition (\ref{3r6UfT}) then determines the Lagrange multiplier as
\begin{equation}
    \mu
    =
    \sqrt{\frac{9 V (W-{\varrho g}\Omega)}{8 \pi \eta \ell}} ,
\end{equation}
whence (\ref{hPQPpg}) and (\ref{23HK55}) become
\begin{equation}\label{2vAUVN}
    h(s)
    =
    \sqrt{\frac{2 \pi \eta \ell V}{W-{\varrho g}\Omega}}
    \frac{\rho(s)}{z'(s)} ,
\end{equation}
and
\begin{equation}\label{2fxa4n}
    D(V)
    =
    \sqrt{\frac{ V (W-{\varrho g}\Omega)^3}{2 \pi \eta \ell}}
\end{equation}
respectively.

For a spherical probe of radius $a$, we have
\begin{equation}
    \rho(s) = a \sin(s/a),
    \qquad
    z'(s) = \sin(s/a) ,
\end{equation}
with $s$ measured from the bottom of the probe, and
\begin{equation}\label{RZAt99}
    \ell
    =
    \frac{\sqrt{\pi} \, \Gamma(5/4)}{\Gamma(7/4)} a
    \approx
    1.74804 \, a ,
\end{equation}
which, inserted into (\ref{2vAUVN}) and (\ref{2fxa4n}), fully determines the fluid layer thickness and the viscous dissipation, respectively. We note, in particular that the fluid layer thickness evaluates to
\begin{equation}\label{2vAUVN}
    h(s)
    =
    \sqrt{\frac{2 \pi \eta \ell V}{W-{\varrho g}\Omega}}
    \, a
    =
    \text{constant},
\end{equation}
i.~e., the fluid layer around a spherical probe is of uniform thickness to within the accuracy of the analysis.

\begin{figure}[ht]\centering
   \includegraphics[width=0.75\linewidth]{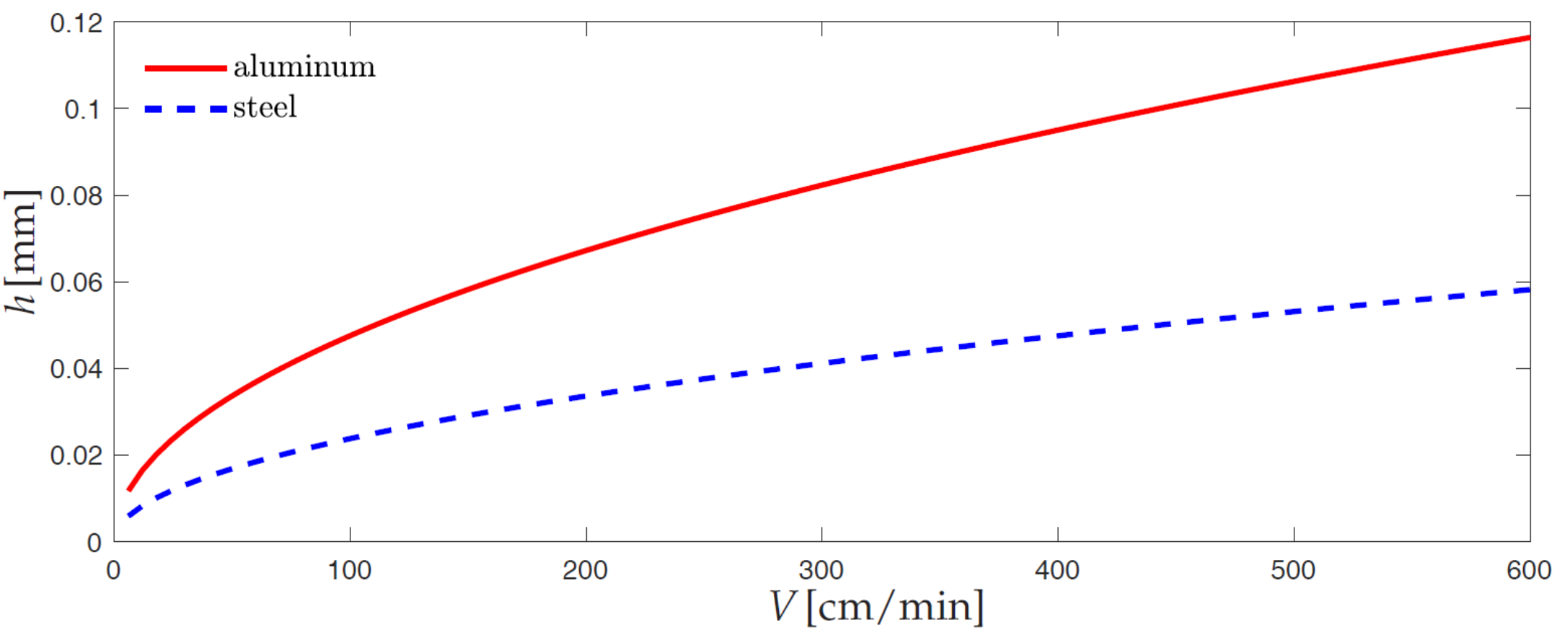}
	\caption{Thickness of the liquid layer surrounding a solid spherical probe as a function of speed of descent for two different weights and under the conditions of the supporting experiment.} \label{fig:liquidlayer}
\end{figure}

The dependence of the liquid layer thickness on the speed of descent for the conditions of the supporting experiment are shown in Fig.~\ref{fig:liquidlayer}. In all cases, the thickness of the layer is exceedingly small, which justifies {\sl a posteriori} the lubrication approximation adopted in the analysis. As may be seen from the figure, the thickness increases with speed and decreases with the weight of the probe.

We note that, in (\ref{2vAUVN}) and (\ref{2fxa4n}), the velocity of descent $V$ is itself a function of the reduced weight $W-{\varrho g}\Omega$ and other factors. This dependence is determined subsequently by recourse to energy balance.

\section{Speed of descent}

\smallskip

We assume that the speed of descent of the probe is the result of a balance between heat supply, radiative heat losses, viscous dissipation, latent heat and gravity in a control volume coincident with the liquid layer. From the previous analyses, this balance of energy takes the expression
\begin{equation}\label{13wqgD}
    \dot{Q}_0
    +
    (W - \varrho g \Omega) V
    =
    D(V)
    +
    \dot{Q}_\infty(V)
    +
    L A V
\end{equation}
where $\dot{Q}_0$ is the power supply, $W$ is the weight of the probe, $\varrho$ is the mass density of ice, $g$ is the acceleration of gravity, $\Omega$ is the volume of the probe, $V$ is the speed of descent, $D$ is the viscous dissipation, $\dot{Q}_\infty(V)$ is the radiated heat rate, $L$ is the latent heat rate per unit volume required to melt the solid ice and $A$ is the cross-sectional area of the probe. Equation (\ref{13wqgD}) supplies an equation that can be solved for the velocity of descent $V$.

Equation (\ref{13wqgD}) simply requires that the power supplied by the probe and the gravitational power be exactly spent on viscous dissipation, radiation losses and latent heat of melting. We note that, although the liquid water refreezes behind the probe, we assume that the latent heat of melting is not recovered by the liquid layer. This assumption is justified in view of the larger thermal conductivity of ice compared to that of water, cf.~Table~\ref{tab:material}. Under these conditions, most of the recovered latent heat behind the probe leaks into the ice and does not contribute to the balance of energy of the liquid layer.

\begin{figure}[ht]\centering
\includegraphics[width=0.75\linewidth]{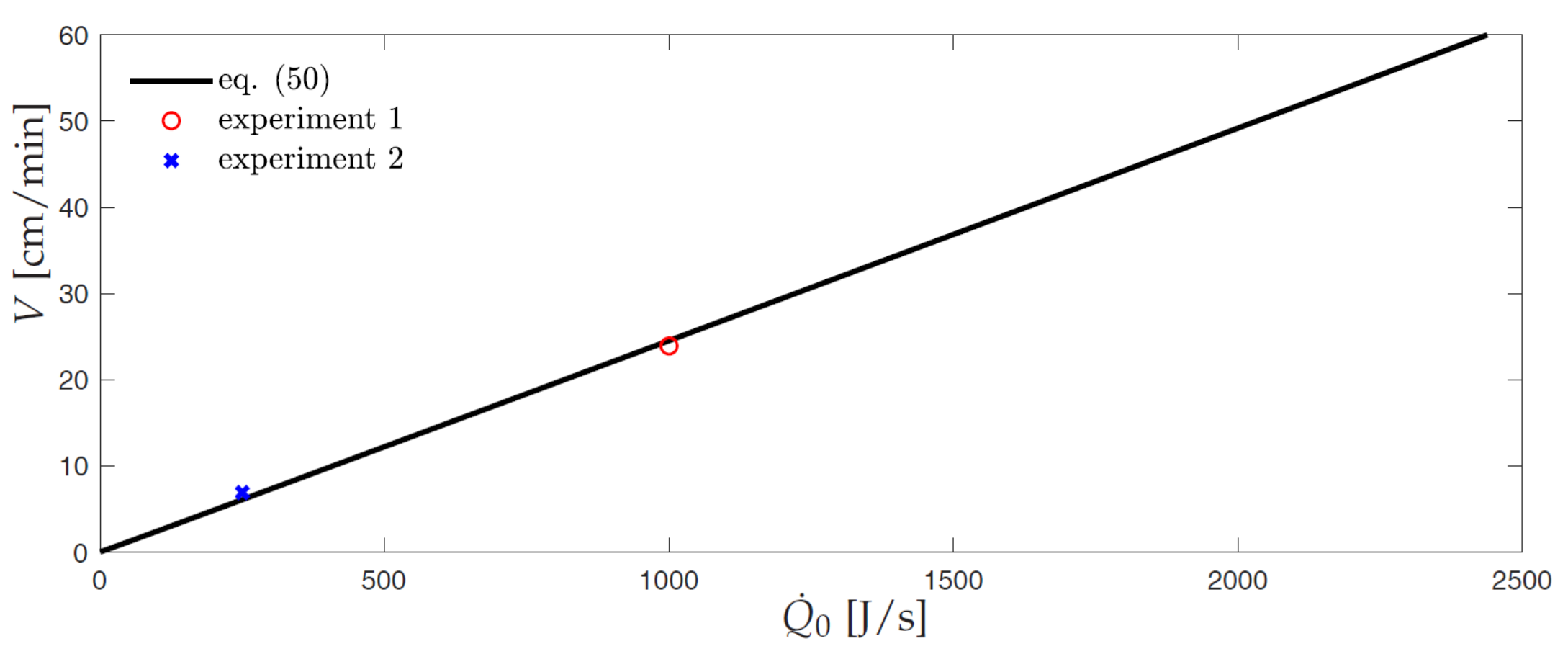}
    \caption{Speed of descent $V$ as a function of power supply $\dot{Q}_0$ for the conditions of the supporting experiment.} \label{fig:speed_heat}
\end{figure}

For a spherical probe, using (\ref{9iekiE}), (\ref{2fxa4n}) and (\ref{f1yV3X}, eq.~(\ref{13wqgD}) specializes to
\begin{equation}\label{7s8EZv}
\begin{split}
    &
    \dot{Q}_0
    +
    \Big( W - \varrho g \Omega \Big) V
    = 
    \sqrt{\frac{ V (W-{\varrho g}\Omega)^3}{2 \pi \eta \ell}}
    +
    2 \pi c V a^2 \,
    (T_m - T_\infty)
    (1 + \coth(\frac{c V a}{2 {\kappa}}))
    +
    \pi a^2 L V ,
\end{split}
\end{equation}
with $\ell$ as in (\ref{RZAt99}) and $\Omega = 4\pi a^3/3$, to be solved for $V$.
For water ice, the viscosity is relatively low and the radiative losses dominate in general. For example, for the values in Table~\ref{tab:material} and the conditions of the supporting experiment, we have: i) $4 \pi a^2 c (T_m - T_\infty) \sim 7.4 \times 10^5$ N, and ii) $A L \sim 16.8 \times 10^5$ N, which overwhelm the remaining terms. The dependence of the resulting speed of descent on power supply is shown in Fig.~\ref{fig:speed_heat}. Also shown in the figure are two corresponding measured data points. Despite the paucity of experimental data, the agreement between theory and experiment is remarkable.

\section{Concluding remarks}

\smallskip

In summary, an energy-balance argument determines the speed of descent to be the result of: i) The power supplied by the probe; ii) the buoyancy and weight of the probe; iii) viscous dissipation in the liquid layer surrounding the probe; iv) radiative heat losses, and v) latent heat of solid ice. The present analysis differs from the conventional model based on latent heat in that it additionally accounts for radiation losses, the effect of the viscous layer of liquid water surrounding the probe and the weight of the probe. In particular, the thickness of the liquid layer is unknown at the outset and its determination is part of the analysis, in the spirit of Stefan problems. Remarkably, due to the thinness of the liquid layer the Stefan problem can be solved analytically by recourse to a lubrication approximation. For water ice, the viscosity correction is small and radiative losses and latent heat dominate. The agreement between the velocity of descent predicted by the theory as a function of power supply and experimental measurements is quite remarkable and provides a modicum of validation of the theory.

\section{Acknowledgment}

\smallskip

We thank the PEP-120217 student group of the University of Siegen, Germany, under the guidance of R.~N\"otzel, for their participation in the experimental part of the study.

\bibliographystyle{asmems4}
\bibliography{lit_europa}

\end{document}